# Topological photonic cavities based on dissimilar Bragg gratings

Alejandro Sánchez-Sánchez[1,*], José Manuel Luque-González[1], Gauthier Krizman[2], Dorian Oser[3], Paula Nuño Ruano[2], David González-Andrade[1], David Medina Quiroz[2], Samson Edmond[2], Alejandro Ortega-Moñux[1], Jens H. Schmid[3], Pavel Cheben[3], Laurent Vivien[2], Iñigo Molina-Fernández[1], J. Gonzalo Wangüemert-Pérez[1], and Carlos Alonso-Ramos[2]
[1] Telecommunication Research Institute (TELMA), Universidad de Málaga, CEI Andalucía TECH, E.T.S.I. Telecomunicación, 29010 Málaga, Spain
[2] Centre de Nanosciences et de Nanotechnologies, CNRS, Université Paris-Saclay, Palaiseau 91120, France
[3] Institute of Photonics and Advanced Sensing (IPAS), School of Electrical and Mechanical Engineering, Adelaide University, Adelaide SA 5005, Australia
[4] National Research Council Canada, 1200 Montreal Road, Bldg. M50, Ottawa K1A 0R6, Canada
* as.sanchez@uma.es

## Abstract

Topological photonic cavities offer a powerful route to robust optical confinement and enhanced light–matter interactions. Interface states that emerge at the boundary between one-dimensional periodic structures with distinct topological phases, enable cavities with ultra-small mode volumes and intrinsic protection against disorder. Existing implementations typically create the trivial and topological phases by redefining the unit cell on either side of the cavity, so that both periodic structures share the same band structure. This constraint limits design flexibility and the range of accessible devices. Here we introduce a fundamentally different strategy for realizing topological cavities based on combining periodic waveguides with distinct band structures. By exploiting bandgap closing and band inversion in Bragg gratings, we independently control the topological phase and bandgap width of each structure. We experimentally realize silicon topological cavities formed by two different Bragg gratings without period shifting, and observe topological modes despite significant differences between the two gratings. Our results establish a new route to topological photonic cavity design, demonstrating that band inversion between dissimilar Bragg gratings enables cavity formation beyond symmetric constraints and provides a mechanism to engineer optical confinement via mirror asymmetry.

## Introduction

Topological photonics is a field that builds on concepts originally developed in condensed matter physics, where a system's behavior is governed by global topological properties rather than local perturbations. Early demonstrations of topological protection in electromagnetic systems were realized using gyromagnetic photonic crystals operating at microwave frequencies[1]. Since then, topology has emerged as a powerful framework for controlling light, enabling photonic states with unusual robustness and new degrees of freedom for manipulating optical modes[2–13]. Among these developments, topological photonic cavities have attracted particular interest because they confine light through zero-dimensional interface states that arise at the boundary between one-dimensional photonic crystals with distinct topological phases[3,14,15]. These edge states enable optical modes with ultra-small mode volumes while providing intrinsic robustness against fabrication disorder, making them promising for applications such as nanolasers[7,12,16] and optical filtering[17,18]. State-of-the-art topological cavities typically generate the required trivial and topological phases by redefining the unit cell of the photonic crystal on the two sides of the cavity[7,12,16–19]. In this approach, the two mirrors correspond to different choices of unit-cell origin within an otherwise identical periodic structure. Although this strategy guarantees the formation of a topological interface state, it also forces both mirrors to share the same band structure. As a consequence, the confined optical mode penetrates symmetrically into the two mirrors, restricting opportunities for engineering the spatial distribution of the optical field. Independent control of modal penetration is advantageous in devices combining spatially separated gain and loss, parity–time-symmetric photonic systems[20], or structures incorporating regions with different absorption levels[21].

Here we introduce a fundamentally different strategy for realizing one-dimensional topological cavities by combining photonic crystals with dissimilar band structures. Our approach exploits bandgap closure and band inversion in Bragg gratings to independently control the topological phase and the spectral bandwidth of the mirrors. Although the strength of a Bragg grating is commonly assumed to increase monotonically with the corrugation amplitude, under specific conditions the bandgap can close after initially opening at large corrugation strengths[22,23]. Importantly, gratings on the two sides of this bandgap closure possess different topological phases[24]. By harnessing this effect, we design mirrors that support topological interface states while exhibiting distinct band structures. Optomechanical topological cavities based on multilayer Bragg mirrors with different geometries have previously been reported[19]. However, these systems operate at high-order Bragg resonances, which would introduce prohibitively large optical losses in integrated photonic platforms, and they consider only configurations in which both mirrors exhibit nearly identical band structures.

Using our approach, we realize silicon topological cavities formed by two distinct Bragg gratings without relying on unit-cell shifting. We experimentally demonstrate the formation of topological interface modes when combining periodic structures with both nearly identical and strongly different bandgap widths, differing by up to a factor of two. These results establish a new route for the design of topological photonic cavities and substantially expand the available design space for engineering confined optical modes.

## Results

### From Fabry-Pérot cavities to topological cavities

The resonance condition of an optical cavity formed by two mirrors can be written as

$$2\beta L + \phi_{\mathrm{L}} + \phi_{\mathrm{R}} = 2m\pi$$

where $\beta$ is the propagation constant of the guided mode, $L$ is the length of the intracavity region, and $\phi_{(\mathrm{L,R})}$ denote the reflection phases of the left and right mirrors, respectively. The resonance therefore depends not only on the optical path within the cavity but also on the phase accumulated upon reflection at the two mirrors.

For one-dimensional photonic crystals with inversion symmetry, the reflection phase varies continuously across the bandgap in a manner determined by the topological phase of the lattice[24]. In structures with a trivial phase, the reflection phase evolves from 0 to -π across the bandgap, whereas in structures with a topological phase it evolves from π to 0. Conventional Fabry–Pérot cavities combine two mirrors with the same type of phase response, typically trivial mirrors, and achieve the resonance condition by adjusting the physical cavity length, $L$. Topological cavities follow a different principle. They combine mirrors with opposite topological phases, so that the sum of the reflection phases crosses zero at the center of the bandgap. As a result, the resonance condition can be satisfied even in the absence of an intracavity region ($L$ = 0), leading to a localized optical mode confined at the interface between the two photonic crystals.

The proposed topological cavity is illustrated in Fig. 1. The structure combines two rectangular Bragg gratings with different corrugation amplitudes, which harness bandgap closure and subsequent band inversion. The unit cell of the grating consists of a structure with average width $w_a$, comprising a periodic rectangular corrugation of amplitude $w_c$, as shown in Fig. 1. This mirror geometry preserves inversion symmetry. The Bragg grating on one side of the cavity is designed with a corrugation width below the bandgap-closure point, corresponding to a trivial phase, whereas the grating on the opposite side has a corrugation width above the closure point and therefore exhibits a topological phase. The cavity is designed on a 220-nm-thick silicon-on-insulator platform with air cladding, considering only the fundamental transverse-electric mode. The average waveguide width is chosen as $w_a$ = 400 nm to ensure single-mode operation. A grating period of Λ = 375 nm centers the first-order Bragg bandgap at the 1550 nm telecom wavelength. The duty cycle is fixed at 80%, which helps maintain the bandgap centered near the target wavelength as the corrugation amplitude increases.

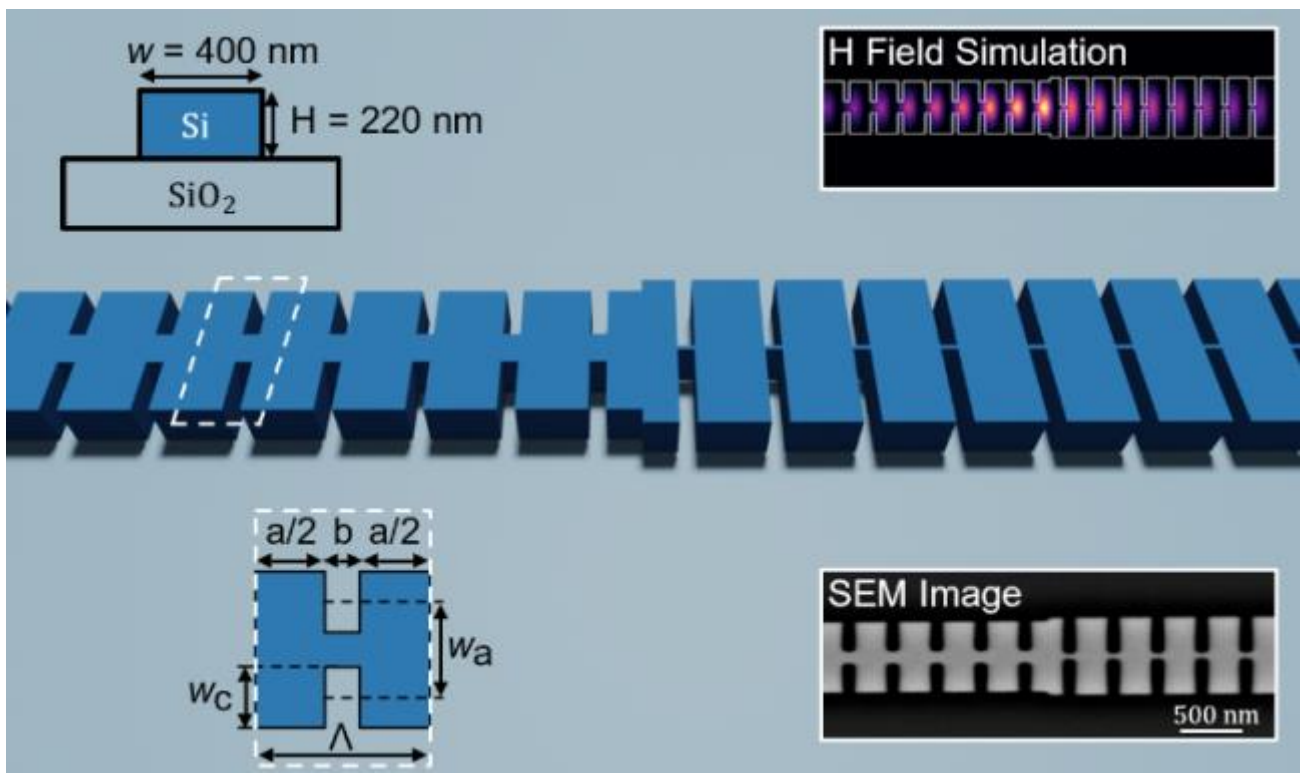

**Fig. 1 | Topological cavity based on dissimilar Bragg gratings.** Schematic of the proposed structure, which combines two rectangular Bragg gratings with different corrugation amplitudes to exploit bandgap closure and subsequent band inversion. Insets show a cross-section of the device, a Fourier-type eigenmode expansion simulation of the confined topological mode, and a scanning electron microscope (SEM) image of a fabricated cavity.

## Band inversion and topological cavity formation in Bragg gratings

Figures 2a–c shows the simulated band structures (see Methods) for three representative corrugation widths: before the bandgap closure, at the closure point (Dirac point), and after

the closure. As illustrated in the insets, which represent the real part of the $E_y$ field component, the Bloch modes at the band edges exhibit opposite symmetries before and after the bandgap closure, indicating band inversion and therefore a change in the topological phase. Figure 2d summarizes the evolution of the upper and lower band-edge wavelengths of the first-order bandgap as a function of the corrugation width. The bandgap closes at $w_c \approx 310$ nm, separating two regions with opposite topology: structures with $w_c$ <310 nm exhibits a trivial phase, whereas structures with $w_c$ >310 nm exhibit a topological phase.

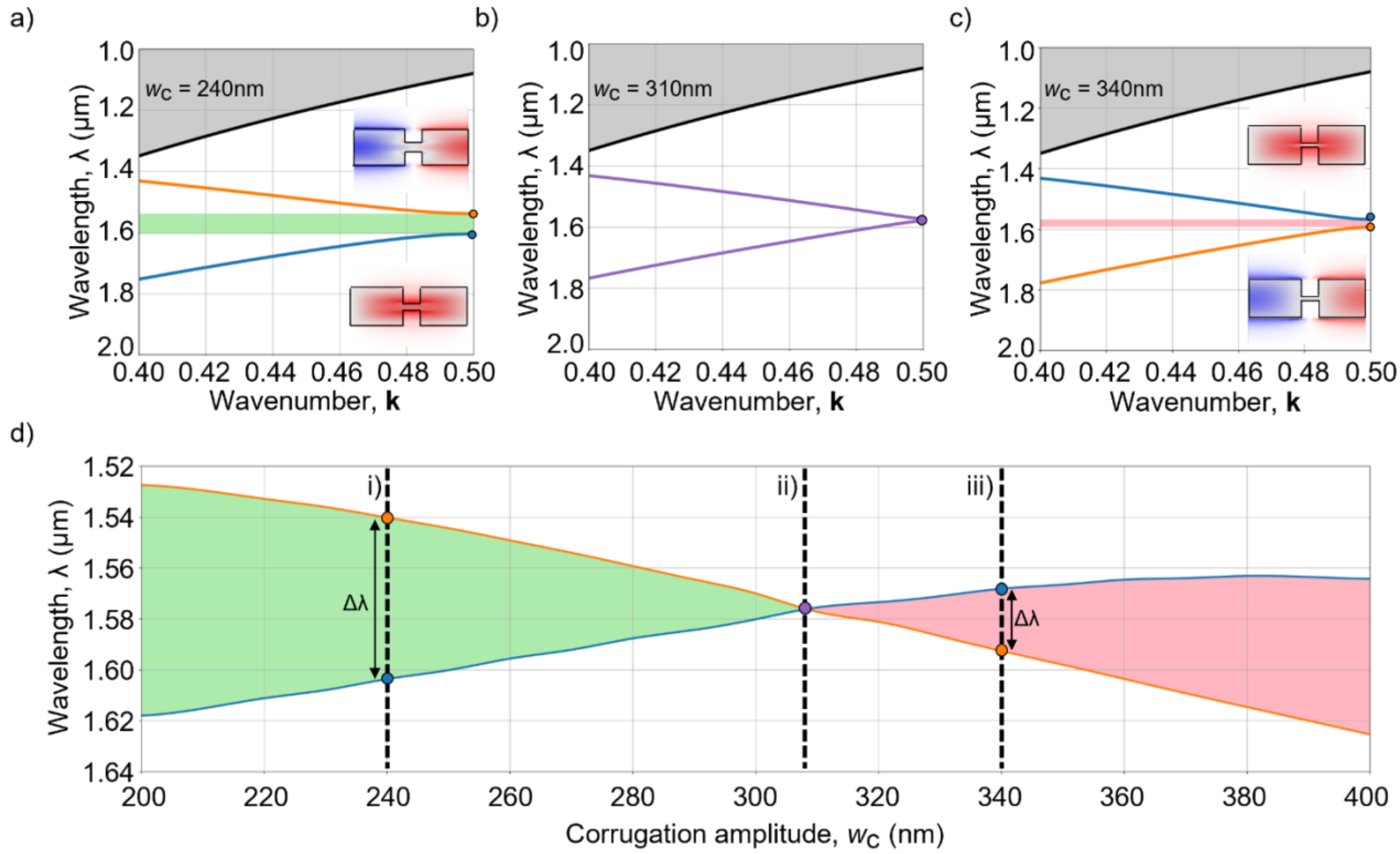

**Fig. 2 | Band inversion and topological phase transition**. a–c, Calculated band structures for corrugation widths $w_c$ = 240 nm, 310 nm and 340 nm. The bandgap progressively narrows and closes near $w_c$ = 310 nm, where the two bands cross. Insets show the spatial field distributions of the Bloch modes at the band edges, which exchange symmetry across the transition, indicating band inversion. d, Wavelengths of the upper and lower edges of the first-order bandgap as a function of the corrugation width $w_c$. The crossing point, corresponding to a Dirac point, marks the transition between two topological phases of the structure.

We further analyze the Bragg gratings and the resulting topological cavity using 3D finite-difference time-domain (FDTD) simulations (see Methods). FDTD is employed both to extract the reflection-phase response of individual gratings, providing a direct probe of their topological character, and to simulate the full cavity formed by interfacing two gratings, enabling direct evaluation of the resonance condition. Two configurations are considered: the first consists of gratings with similar grating strengths (comparable bandwidths), with corrugation widths $w_c$ = 285 nm and $w_c$ = 340 nm, representing a symmetric configuration analogous to the conventional topological cavity implementations. In the second configuration, gratings with dissimilar grating strengths are combined, with $w_c$ = 240 nm and $w_c$ = 340 nm, leading to markedly different bandgap widths. In both cases, the gratings are designed to lie on opposite sides of the Dirac point, resulting in opposite topological phases.

We also include a third grating, with $w_c$ = 305 nm, corresponding to the Dirac point, as identified by the closure of the bandgap. The Dirac-point position differs by approximately 3 nm from the band-structure calculations, which we attribute to differences between simulation methods.

Figure 3a,b shows the transmission spectra of uniform gratings composed of 100 periods for both configurations. The symmetric case exhibits comparable bandgap widths for the two gratings, whereas the asymmetric case shows significantly different bandwidths. In both cases, the bandgap centers remain approximately aligned due to the fixed duty cycle, ensuring that both gratings exhibit high reflectivity over a common spectral region. Figure 3c,d shows the reflection phase across the bandgap for the gratings forming the cavity (see Methods). As discussed in the previous section, the phase of the reflection coefficient is directly related to the topological phase of the structure. Accordingly, the gratings with corrugation width below the Dirac point exhibits a reflection phase evolving from 0 to $-\pi$, corresponding to a trivial phase, whereas the grating above the Dirac point exhibits a phase evolution from $\pi$ to 0, corresponding to a topological phase. This change of sign is a direct consequence of the topological transition. This behavior reflects the band inversion discussed in Figure. 2d. The sum of the reflection phases is also shown, crossing zero near the center of the bandgap and therefore satisfying the resonance condition in Eq. (1). Figure 3e,f shows the corresponding cavity transmission spectra, where a well-defined resonance is observed near 1580 nm in both configurations, in agreement with the phase accumulation condition.

The results demonstrate that bandgap closure in these Bragg gratings originates from a band inversion, which corresponds to a topological transition manifested in the reflection phase and leading to cavity formation. While the symmetric configuration is analogous to conventional topological cavity implementations based on Bragg mirrors sharing the same band structure, the asymmetric case shows that cavity formation is not restricted to this condition. Instead, it is governed by the phase-matching condition between mirrors with opposite topological phases, provided that their bandgaps overlap sufficiently to enable phase cancellation. This remains valid both for comparable and strongly dissimilar grating strengths, despite significant differences in bandgap widths. It confirms the robustness of the proposed Bragg-assisted topological cavity while opening additional degrees of freedom to engineer the confinement of the cavity mode through independent control of the mirror strengths.

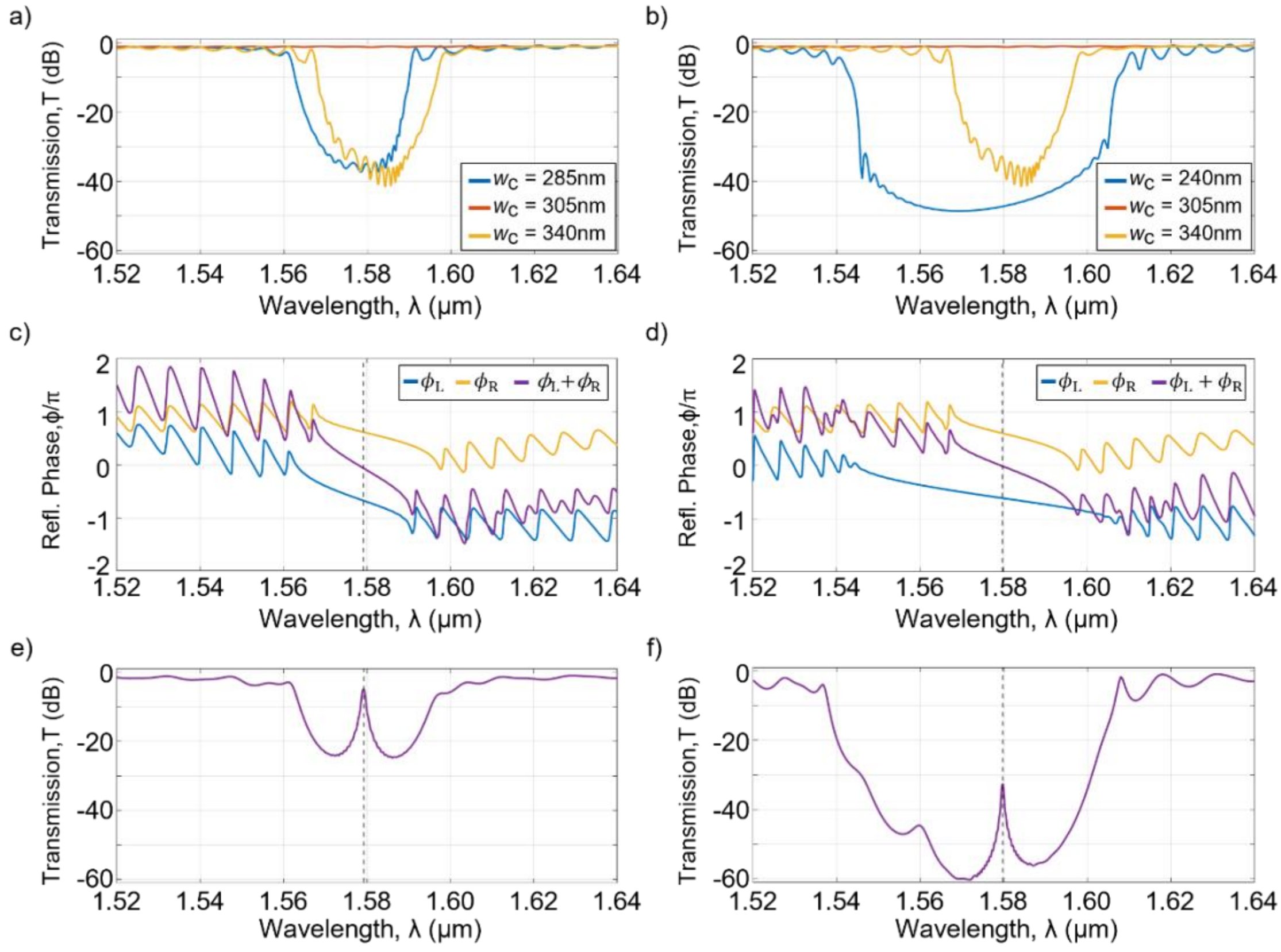

**Fig. 3 | 3D-FDTD simulation of Bragg-assisted topological cavity**. a,b, Uniform Bragg grating with 100 periods and corrugation of 240 nm, 285 nm, 304 nm, 340 nm. c,d, Phase of the reflection coefficient for representative gratings forming the cavity, corresponding to a configuration with similar grating strengths (285 nm and 340 nm) and a strongly asymmetric configuration (240 nm and 340 nm), respectively. e,f, Transmission spectra of the Bragg-assisted topological cavity, each formed by two gratings with 50 periods, for the similar and dissimilar configuration.

## Experimental observation of bandgap closure in Bragg gratings

To experimentally probe the bandgap closure, a series of uniform Bragg gratings was fabricated (see Methods) each consisting of 100 periods with corrugation widths ranging from 120 nm to 380 nm. Figure 4a shows the measured transmission spectra for three representative corrugation widths: 300 nm, 320 nm and 340 nm. For corrugation widths of 300 nm and 340 nm the gratings exhibit similar bandgaps of approximately 25 nm, whereas at 320 nm the bandgap closes completely. This closure corresponds to the formation of a Dirac point, providing the condition required to realize the proposed topological cavities. Figure 4b summarizes the experimental results by plotting the wavelengths of the upper and lower band edges as a function of the corrugation width. The two band edges approach and cross at the Dirac point, confirming the experimentally observed bandgap closure. Notably, the Bragg center wavelength remains nearly constant across the full range of corrugation amplitudes. This behavior arises from the definition of the corrugation geometry together with the optimized duty cycle, which stabilizes the Bragg condition as the corrugation amplitude varies.

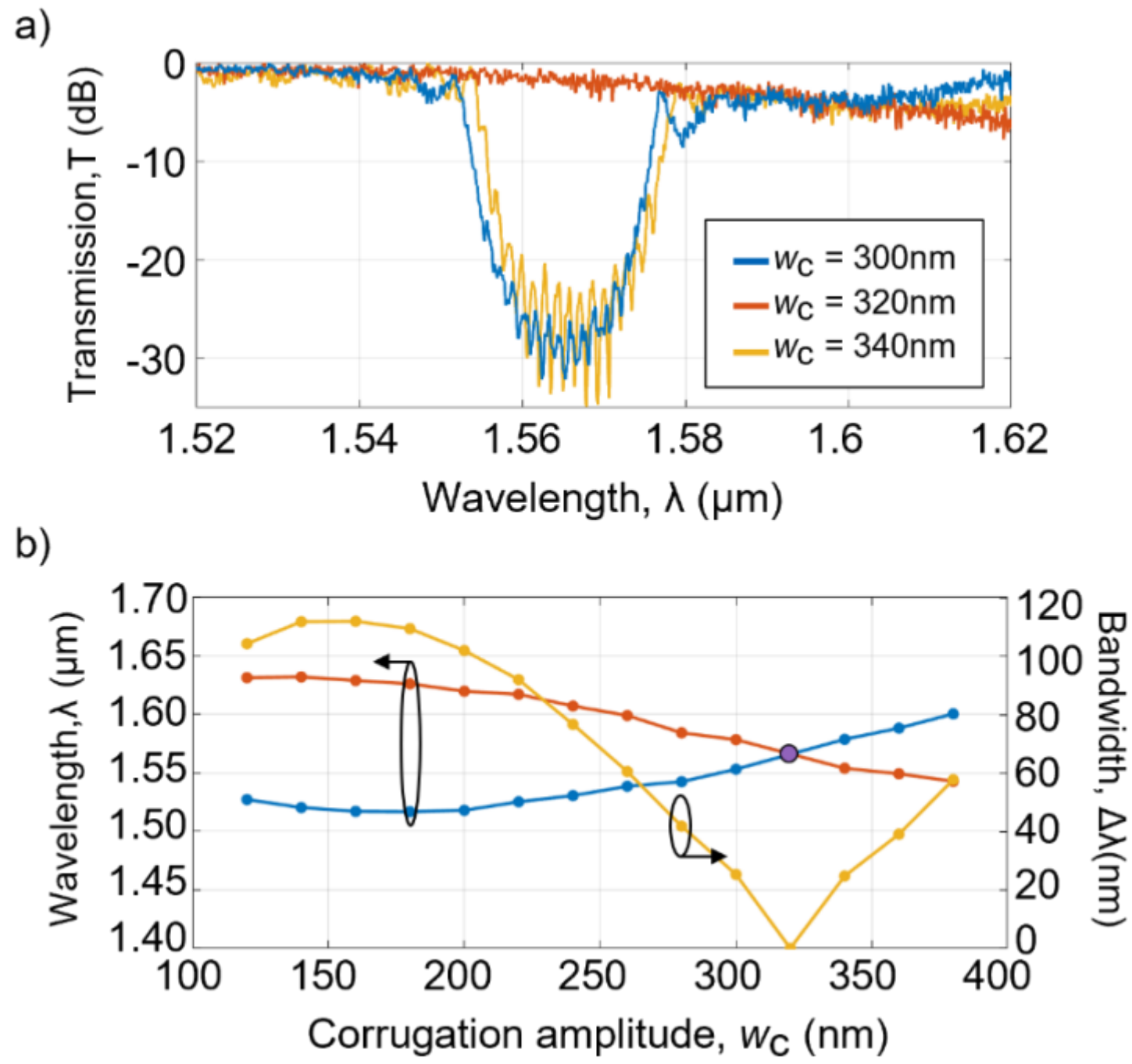

**Fig. 4 | Experimental demonstration of band closure.** a, Transmission spectrum measurement of the Bragg gratings with various corrugation amplitudes. b, Wavelengths of the upper and lower edges and bandwidth measured as a function of the Bragg gratings corrugation amplitudes.

## Topological cavity formation in dissimilar Bragg gratings

To investigate the emergence of topological cavity modes, we fabricated a series of devices combining Bragg gratings with different corrugation widths on the two sides of the cavity, each grating consisting of 50 periods. The gratings were chosen to lie on opposite and same sides of the experimentally identified Dirac point. The measured transmission spectra for the different combinations of corrugation amplitudes are shown in Fig. 5a. Cases in which the two gratings have identical corrugation widths are highlighted in blue. As expected, these spectra reveal the progressive opening and closing of the bandgap with increasing corrugation amplitude, with the Dirac point occurring at $w_c$ = 320 nm (shaded in purple). Combinations in which both gratings lie on the same side of the Dirac point are highlighted in red (two topological phases) and yellow (two trivial phases). In these configurations the two mirrors share the same topological phase, and no localized interface state is formed, resulting in the absence of a cavity resonance. By contrast, structures combining gratings with opposite topological phases are highlighted in green. In these cases, a zero-dimensional edge state emerges at the interface between the two gratings, forming a topological cavity. A cavity resonance is observed not only when the two gratings exhibit similar bandgap widths—for example $w_{c,l}$ = 280 nm and $w_{c,r}$ = 380 nm - but also when the bandwidths differ significantly, by nearly a factor of two.

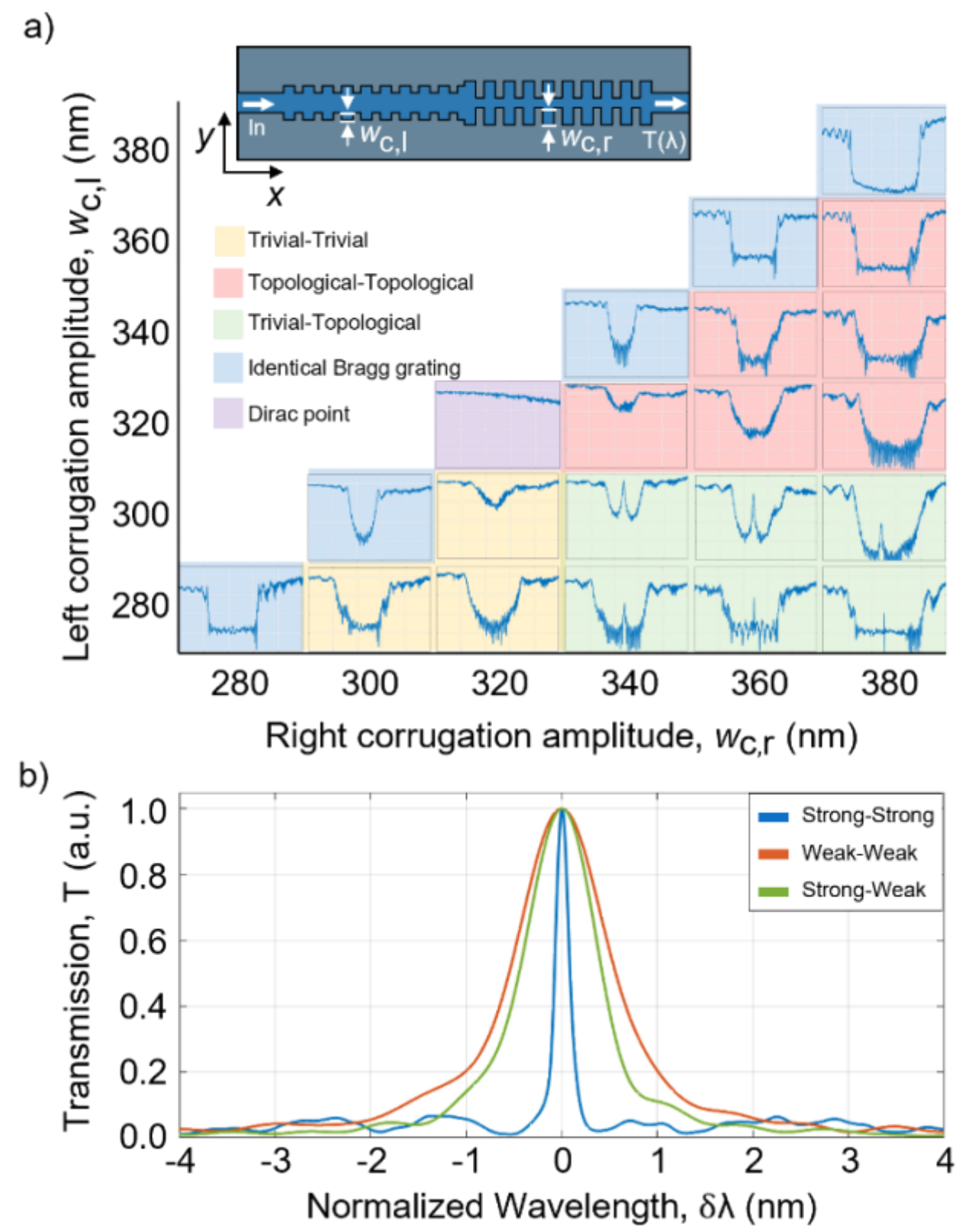

**Fig. 5 | Experimental demonstration of topological cavities formed by dissimilar Bragg gratings.** a, Measured transmission spectra for cavities formed by different combinations of corrugation widths in the left and right gratings. A localized resonance is observed only when the two gratings have opposite topological phases, confirming the formation of a topological interface state. The cavity mode persists for both similar and strongly dissimilar bandgap widths. Inset: schematic of the cavity formed by interfacing two rectangular Bragg gratings with different corrugation widths. b, Zoomed-in spectra for representative configurations with different relative grating strengths, showing the cavity resonances used to extract the quality factors.

A closer inspection of representative topological cavity configurations is provided in Fig. 5b, where we show zoomed-in transmission spectra, after removing residual Fabry–Pérot oscillations from the input and output gratings (see Methods). Three cases are considered, corresponding to different relative grating strengths: i) a cavity formed by two strong mirrors with a bandwidth of ≈ 40 nm ($w_{c,l}$=280 nm and $w_{c,r}$=360 nm), ii) a cavity comprising two weaker mirrors with reduced bandwidth of ≈ 20 nm ($w_{c,l}$=300 nm and $w_{c,r}$=340 nm), and iii) a strongly asymmetric configuration combining a strong and a weak mirror ($w_{c,l}$=300 nm and $w_{c,r}$=360 nm). The resonances exhibit markedly different linewidths, due to variations in the effective reflectivity of the cavity mirrors across the three configurations. The quality factors, extracted from Lorentzian fits to the resonances, are Q ≈ $1.1 \times 10^4$, $1.5 \times 10^3$, and $2 \times 10^3$ for the strong–strong, weak–weak, and strong–weak configurations, respectively. These values correspond to loaded quality factors, currently limited by coupling to the input and output waveguides; increasing the grating length would reduce these external losses and allow the system to approach its intrinsic quality factor, which is ultimately limited by radiation losses associated with the cavity mode profile. Importantly, the asymmetric configuration does not lead to a degradation of performance compared to the symmetric cases with similar mirror strength. Instead, the quality factor is primarily determined by the effective reflectivity of the mirrors, rather than by whether the configuration is symmetric or asymmetric. These results demonstrate the additional design flexibility enabled by asymmetric configurations, allowing tailoring of the cavity mode confinement by independently control of the two mirrors.

## Discussion

In summary, we have proposed and experimentally demonstrated a new approach to implement one-dimensional topological photonic cavities based on the combination of Bragg gratings with dissimilar band structures. By exploiting bandgap closure and band inversion in rectangular gratings, we independently control the topological phase and spectral properties of the mirrors, enabling the formation of zero-dimensional interface states without relying on conventional unit-cell shifting.

Using silicon photonic structures operating at telecommunication wavelengths, we experimentally observe topological cavity modes formed at the interface between gratings with opposite topological phases. Importantly, these modes persist even when the two gratings exhibit substantially different bandgap widths, differing by nearly a factor of two. This behavior contrasts with conventional implementations of topological cavities, in which the two mirrors necessarily share identical band structures[7,17,18,25].

Our measurements also provide direct experimental evidence of the bandgap closure that defines the Dirac point separating the two topological phases. While signatures of bandgap closure have been reported previously in photonic systems[22], experimental evidence of the associated band inversion has remained elusive and has largely relied on comparisons between simulations and measurements of bandgap widths or resonance positions in Bragg-loaded resonators[23]. Here, the observation of localized interface modes between gratings on opposite sides of the Dirac point provides direct experimental evidence of band inversion. Indeed, the formation of a topological cavity requires the two mirrors to exhibit opposite topological phases, which can only arise through band inversion across the Dirac point.

The ability to form topological cavities from dissimilar periodic structures significantly expands the design space of topological photonic devices. In particular, independent control of the mirror band structures enables new opportunities for engineering asymmetric modal confinement, which could be exploited in systems combining spatially separated gain and loss, parity–time-symmetric photonic platforms, or devices incorporating regions with different doping levels or absorption.

More broadly, the concept demonstrated here shows that topological confinement does not require identical photonic lattices but can emerge from interfaces between distinct periodic structures with appropriately engineered band inversion. This additional flexibility may facilitate the implementation of topological cavities across a wide range of nanophotonic platforms and functionalities, including integrated lasers, nonlinear photonic devices and quantum light sources.

## Methods

### Photonic band structure calculations

Band-structure calculations were performed using the MIT Photonic Bands (MPB). The unit cell consists of a 220-nm-thick silicon core on a silicon dioxide ($SiO_2$) substrate with air as the superstrate medium. The refractive indices of silicon and silica were taken as 3.476 and 1.444, respectively, at 1550 nm. The grating period was set to Λ=375 nm to position the Bragg wavelength at this wavelength. The grating geometry is defined by a duty cycle of 80% and an average waveguide width of 400 nm, while the corrugation width $w_c$ is varied. The in-plane transverse electric (TE) polarization is selected by exploiting the odd symmetry of the electric field along the transverse y direction. The band structure is obtained by sweeping the normalized Bloch wavevector across the Brillouin zone (Fig. 2a–c), or, alternatively, at the Brillouin zone edge ($\mathbf{k}=\pi/\Lambda$) and sweeping $w_c$ to track the evolution of the band edges (Fig. 2d).

## Finite-difference time-domain simulations and phase extraction

Three-dimensional finite-difference time-domain (FDTD) simulations were carried out using Tidy3D. The full device geometry was included, comprising the Bragg gratings or topological cavity, with a silicon core on a buried oxide layer and air as the superstrate medium, as seen in Fig. 6a. Perfectly matched layers (PML) were used as boundary conditions in all directions. A uniform spatial mesh of 20 nm was employed. The structures were excited using a fundamental TE mode source. Reflection and transmission spectra were obtained using frequency-domain monitors placed 1 µm away from the input and output grating interfaces, respectively.

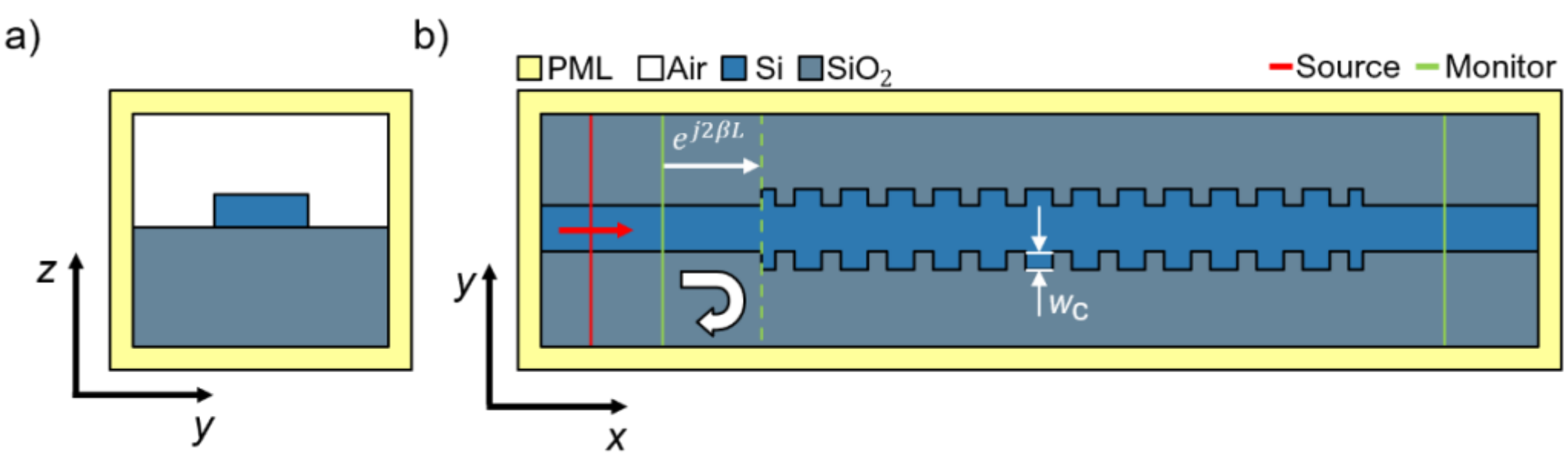

**Fig. 6 | FDTD schematic. a,** Cross-section of the simulated structure, consisting of a silicon core on a $SiO_2$ substrate with air as the superstrate. **b,** Schematic of the simulation setup, showing the excitation source and the reflection monitor. The reflection phase is referenced to the grating input plane by applying a phase correction factor.

The reflection phase was extracted by first recording the reflected field at the monitor plane and then referencing it to to the grating interface (Fig. 6b). This was achieved by applying a phase correction factor $e^{j2\beta L}$, where L is the distance between the monitor and the grating and β is the propagation constant of the guided mode. The propagation constant β was obtained from independent modal simulations of the waveguide over the wavelength range 1.52-1.64 µm. This procedure removes the propagation phase accumulated between the monitor and the interface, yielding the intrinsic reflection phase of the grating. Since the reflection phase is defined modulo 2π, a continuous branch is obtained by phase unwrapping to ensure a smooth spectral evolution within the bandgap.

## Device fabrication

The devices were fabricated on a 220-nm-thick silicon-on-insulator (SOI) platform with a 3-µm buried oxide ($SiO_2$) layer at the Centre de Nanosciences et de Nanotechnologies (C2N) . The photonic structures were defined by electron-beam lithography, followed by inductively coupled plasma reactive-ion etching (ICP–RIE) to transfer the patterns into the silicon layer. After fabrication, the structures were left exposed to air above the silicon layer.

## Device measurements

Optical characterization was performed using a tunable laser source (Santec TSL-770) together with a polarization controller to align the input polarization for selective excitation of the fundamental on-chip TE mode. Light was coupled into and out of the chip through focusing grating couplers using standard single-mode fibers (SMF-28). Transmission spectra were obtained by sweeping the laser wavelength while recording the output power with an optical power meter (Santec MPM-212). All measurements were referenced to the straight waveguide transmittance.

## Resonance fitting and data analysis

The cavity resonances were analyzed by fitting the measured transmission spectra to a Lorentzian line shape in linear power units. Prior to fitting, Fabry–Pérot oscillations arising from the grating couplers were removed using a minimum-phase filtering approach[26]. Specifically, the spectral response was transformed into the time (optical path length) domain, where unwanted reflections associated with the grating interfaces were identified and filtered, while preserving the temporal window containing the cavity resonance. The processed signal was then transformed back into the spectral domain for analysis.
The resulting resonance peaks were fitted to a Lorentzian function, yielding excellent agreement with the experimental data across all configurations. The quality factor Q was extracted from the fitted resonance linewidths as $Q=\lambda_0/\Delta\lambda_{FWHM}$, where $\lambda_0$ is the resonance wavelength and $\Delta\lambda_{FWHM}$ is the full-width at half-maximum (FWHM) obtained from the fit.

## Funding

This work has been funded by the Agence Nationale de la Recherche (ANR-KASHMIR-22-CE24-0021) and the European Union's Horizon Europe (ERC-SPRING, under the Grant Agreement 101087901, ERC-CRYPTONIT, under the Grant Agreement 101097804, Marie Sklodowska-Curie grant agreement No. 101062518). The fabrication was performed with the C2N micro nanotechnologies platforms and partly supported by the RENATECH network and the General Council of Essonne. Collaborative Science, Technology and Innovation Program (CSTIP) Small Teams (ST-R2-01-02) and High Throughput and Secure Networks Challenge Program at the National Research Council Canada (HTSN 210); Ministerio de Ciencia e Innovación, Spain (PRE2022-000085); Universidad de Málaga.